\documentstyle[preprint,eqsecnum,aps]{revtex}
\begin{document}
\draft
\preprint{Submitted to {\bf Phys. Rev. B}, UW-Madison preprint}
\title{
$Pb_{0.4}Bi_{1.6}Sr_{2}Ca_{1}Cu_{2}O_{8+x}$ and Oxygen Stoichiometry:\\
Structure, Resistivity, Fermi Surface Topology and Normal State Properties}
\author{Jian Ma,$^{a,c}$ P. Alm\'{e}ras,$^b$ R.J. Kelley,$^{a,c}$
H.Berger,$^b$\\
G. Margaritondo,$^b$ X.Y. Cai,$^c$, Y. Feng,$^c$ D. Larbalestier,$^{a,c}$
and M.Onellion$^{a,c}$}
\address{$^{a}$Department of Physics, University of Wisconsin-Madison,
Madison, WI 53706\\
$^{b}$Institut de Physique Appliqu\'{e}e, Ecole Polytechnique F\'{e}d\'{e}rale,
CH-1015 Lausanne, Switzerland\\
$^{c}$Applied Superconductivity Center, University of Wisconsin-Madison,
Madison, WI 53706}
\date{Received\ \ \today}
\maketitle

\begin{abstract}
$Pb_{0.4}Bi_{1.6}Sr_2CaCu_2O_{8+x}$ ($Bi(Pb)-$2212) single crystal samples
were studied using transmission
electron microscopy (TEM),
$ab-$plane ($\rho_{ab}$) and $c-$axis ($\rho_c$)
resistivity, and high resolution angle-resolved ultraviolet photoemission
spectroscopy (ARUPS). TEM reveals that the
modulation in the $b-$axis for $Pb(0.4)-$doped $Bi(Pb)-$2212 is dominantly of
$Pb-$type that is not sensitive to the oxygen content of the system, and the
system clearly shows a structure of orthorhombic symmetry. Oxygen annealed
samples
exhibit a much lower $c-$axis resistivity and a resistivity minimum at
$80-130$K. He-annealed samples exhibit a much higher $c-$axis resistivity and
$d\rho_c/dT<0$ behavior below 300K. The Fermi surface (FS) of oxygen annealed
$Bi(Pb)-$2212 mapped out by ARUPS has a pocket in the FS around the
$\bar{M}$ point and exhibits orthorhombic
symmetry. There are flat, parallel sections of the FS, about 60\% of the
maximum possible along $k_x = k_y$, and about 30\% along $k_x = - k_y$. The
wavevectors connecting the flat sections are about $0.72(\pi, \pi)$ along
$k_x = k_y$, and about $0.80(\pi, \pi)$ along $k_x = - k_y$, rather than
$(\pi,\pi)$. The symmetry of the near-Fermi-energy dispersing states in the
normal state changes between oxygen-annealed and He-annealed samples.
\end{abstract}

\narrowtext
\section{Introduction}
\label{sec:level1}
For cuprate superconductors, the shape of the Fermi surface and the
properties of the normal state electronic states has been of continuing
interest. The shape of the Fermi surface is of interest because it
constrains theoretical models.
In $YBa_2Cu_3O_{7-x}$, excellent agreement has been obtained between
angle-resolved photoemission experiments\cite{Campuzano,Tobin,Liu1,Liu2}
and local density
approximation (LDA) band structure calculations.\cite{Massidda1,Yu,Pickett1}
More recently, K. Gofron
et al.\cite{Gofron} have reported an extended van Hove singularity
for $YBa_2Cu_3O_{7-x}$. A. Abrikosov\cite{Abrikosov} has argued
that such a band structure feature is important in
understanding the high superconducting transition temperature ($T_c$).

The symmetry of electronic states comprising the Fermi surface is
identified in $YBa_2Cu_3O_{7-x}$\cite{Liu1,Liu2} because the material is
underdoped for $6.35 < x < 6.95$ and oxygen is removed predominantly from
the chains.\cite{Jorgensen,Claus,Veal}
The $c-$axis resistivity $\rho_c$ increases as $x$
increases, and the interlayer coupling
weakens. Several reports\cite{Koch} indicate that the chain electronic
states are involved in the superconducting properties.

The situation is less clear in $Bi_2Sr_2CaCu_2O_{8+x}$ ($Bi-$2212). C. Olson
and colleagues observed the Fermi surface in this material that
appears to be consistent with LDA calculations.\cite{Massidda2,Krakauer}
They have also reported the presence of an electron-like pocket
around the $(\pi, 0)$ ($\bar{M}$) point for overdoped samples; the pocket was
not observed after removing oxygen.\cite{Olson1,Wells} These results are in
qualitative
disagreement with a rigid band model. Several other authors\cite{Pham} have,
on a variety of grounds, criticized using a rigid band model. Further, an
extended van Hove singularity along the $\Gamma-\bar{M}-Z$ direction in the
Brillouin zone has been reported.\cite{Ma1,Dessau1} The Fermi surface shape
is not
universally agreed on, although most reports argue that the Fermi surface
exhibits orthorhombic\cite{Ma1,Osterwalder} rather than
tetragonal\cite{Dessau1} symmetry.

$Bi_2Sr_2CaCu_2O_{8+x}$ is more complicated for three primary reasons,
all material-related. The $BiO$ planes do not fit perfectly
above/below the $CuO_2$ planes,
so a buckling in the $\Gamma-Y$ direction of the Brillouin zone
results. Also,
oxygen is removed or added primarily to the $BiO$ double layer.
\cite{Pham}
Since the BiO layers possess electronic states near the Fermi
energy,\cite{Massidda2,Krakauer} the interlayer coupling and $\rho_c$ can be
changed dramatically by varying the oxygen stoichiometry.\cite{Kendziora}
Finally, small differences in cation stoichiometry,
and hence disorder, affects the change of oxygen stoichiometry for the same
annealing conditions.\cite{Pham}

The $Pb-$doped BSCCO has attracted much attention since Sunshine
et al.\cite{Sunshine} found that the substitution of lead can enhance
the superconducting temperature $T_c$ in BSCCO multiphase. It was
then found that the lead substitution has a strong influence on
the incommensurate modulation that exists in the $b-$axis.\cite{Ramesh,%
Eibl,XHChen1,XHChen2,Fukushima,Weber} in both $Bi-$2212 and $Bi-$2223.
It appears that
the lead doping could reduce the structure anisotropy (ratio of
$\rho_c/\rho_{ab}$)\cite{Regi}
which is always a factor complicating the interpretation of the
experimental data.\cite{Weber} It is
generally believed that Pb-doping does not perturb the electronic
state critical to forming the superconducting state in this
system.\cite{Wu}. Therefore the electronic structure of $Bi(Pb)-$2212
should be a proper representation of lead-free $Bi(Pb)-$2212.

We have performed extensive measurements on both oxygen annealed and helium
annealed $Bi(Pb)-$2212 single crystals using Transmission Electron
Microscopy (TEM),
in-plane and $c-$axis resistivity, and Angle-Resolved Photoemission
Spectroscopy (ARUPS).
We have used $Bi(Pb)-$2212 samples due to the
larger variation in oxygen stoichiometry that we have been able to reach
with single crystal samples.

Our data in this paper lead to several conclusions for $Bi(Pb)-$2212:
$Bi(Pb)-$2212 is structurally orthorhombic;
oxygen stoichiometry can be very effective in changing the
$c-$axis resistivity, and quite possibly changing the interlayer
coupling;
the Fermi surface exhibits orthorhombic symmetry, incomplete Fermi surface
nesting, and, for sufficiently overdoped samples, a pocket around
the $\bar{M}$ point
in the Brillouin zone;
the symmetry of the electronic states comprising the Fermi surface
changes with oxygen stoichiometry;
the rigid band model does not apply.

\section{Experimental}

The single crystal samples of $Bi_{1.6}Pb_{0.4}Sr_2CaCu_2O_{8+x}$
($Bi(Pb)-$2212) were
prepared using the method similar to that of Wu et al.\cite{Wu}
The crystals were
characterized using four-point resistivity, ac susceptibility, X-ray
diffraction and transmission electron microscopy (TEM).

Transmission electron microscopy measurements were carried out with
a JEOL 200CX microscope operating at 200 KV. The $Bi(Pb)-$2212 single
crystal samples were cleaved repeatedly and thin flakes were then mounted
on to a single-hole gold grid using M-bond adhesive. In order to examine
the crystal from the [100] direction to study the structural modulation, the
crystals were mounted so that the sample foil surface is perpendicular to the
[100] direction. The samples were then mechanically polished and finally
ion-milled at 4.5 KV.

The ARUPS experiments were performed using the four meter normal incidence
monochromator at the Wisconsin Synchrotron Radiation Center in Stoughton, WI.
The beamline provides highly ($>$ 95\%) linearly polarized light with
the photon electric vector in the horizontal plane and with photon energy
resolution better than 10 meV. The angle-resolved photoemission chamber
includes
a reverse-view low energy electron diffraction (LEED) optics used to orient
the sample {\it in situ} after cleaving. The electron
energy analyzer
is a 50 mm VSW hemispherical analyzer mounted on a two axis goniometer, with
an acceptance full angle of $2^\circ$. The base pressure is
$6 \times 10^{-11}$ torr. The incidence angle between the photon Poynting
vector and surface normal was $45^\circ$ unless otherwise noted.

For photoemission studies,
the samples were transferred from a load lock chamber with a
base pressure of $5 \times 10^{-9}$ torr to the main chamber, and
cleaved at 30K in a vacuum of $6-8 \times 10^{-11}$ torr. The sample holder
includes the capability to rotate the sample {\it in situ} about the surface
normal, at low temperatures, for precision alignment with respect to the
photon electric field. To measure the normal state electronic band structure,
the temperature was raised to 95K, above $T_{c}$ . The stability of
the temperature was $\pm1K$. For this study, the overall
energy resolution employed was 55 meV unless otherwise stated.

For a quasi-two-dimensional system such as $Bi_{2}Sr_{2}Ca_{1}Cu_{2}O_{8+x}$
the initial state of the
electron can be determined by measuring the component of the electron momentum
parallel to the sample surface ({\bf k}$_{/ \! /}$). By measuring the energy
distribution curves (EDC's) for different directions
($\theta$, $\phi$) of the emitted photoelectron relative to the surface
normal, the {\bf k}$_{/ \! /}$ of the
initial state is derived from the relation:
{\bf k}$_{/\! /}=0.512$\AA$^{-1} \sqrt{E_{kin}}
(\sin\theta \cos\phi \hat{k_{x}} + \sin\phi \hat{k_{y}})$,
where $E_{kin}$ is the kinetic energy of measured
photoelectrons in the unit of eV, $\hat{k_{x}}$ and $\hat{k_{y}}$ denote
unit vectors along horizontal and vertical
directions, respectively.
A freshly {\it in situ} deposited gold film was used as a reference to
determine the Fermi energy.

The modified modulation structure in $Bi(Pb)-$2212 due to lead substitution was
confirmed by our {\it in situ} LEED study. At low electron beam energy (27 eV)
the LEED pattern exhibits an
almost $1 \times 1$ structure
for $Bi(Pb)-$2212 whereas it is a $4.6 \times 1$
pattern for lead-free $Bi-$2212. At higher electron beam energies (above 40 eV)
we do observe a modulation for the oxygen-annealed samples that is absent for
the He/Ar-annealed samples. The sharp and intense LEED spots confirmed
the high quality of the cleaved surfaces which is a prerequisite for
performing ARUPS measurements.

\section{Crystal Structural and Transport measurements}
\subsection{Crystal Structure of B\lowercase{i}(P\lowercase{b})-2212}
The chemical composition and the crystal structure of the $Pb-$doped
$Bi-$2212 single crystals were examined by analytical electron microscopy.
Energy dispersive $X-$ray spectroscopy confirmed the $Pb$ was incorporated
into the $Bi-2212$ crystals. A twin-like domain structure was observed
in the $Bi(Pb)$-2212 crystals, which has rarely been seen before in the
lead-free crystals. Fig.\ \ref{EDP} shows the [001](a) and [100](b)
diffraction patterns on the oxygen annealed $Bi(Pb)$-2212 crystals.
Both patterns show the modulated structure in the $Bi$ system which
can be understood in terms of a basic structure (shown as the indexed
strong main spots) plus a superimposed displacement field (shown a those
satellite spots marked by arrows). From the [100] pattern, there is
only one type of modulation which parallels exactly with the
$b-$ direction. This is the typical $Pb-$type modulation, as has been
discussed previously.\cite{Eibl,XHChen1,XHChen2}  However,
the modulation wavelength
along the $b-$ direction is about $13b$ here, much longer than those
reported previously (typically $8-10b$).\cite{Matsui,Fung,CHChen}
Such an increasing of the modulation wavelength may be related to the
$Pb$ content and the oxygenation effect. No $Bi-$type modulation
\cite{Eibl,XHChen1,XHChen2} was
observed in this sample.

In order to identify the structure symmetry of the crystals, we have
performed convergent beam electron diffraction (CBED) measurements.
Care was paid to avoid any imperfect area when selecting regions
to perform CBED measurements, since any defects or artifacts can break
symmetry.\cite{Humphreys} CBED were performed many times at different
locations in the crystals and Fig.\ \ref{CBED} shows a typical
example of the [001] CBED pattern of the oxygen annealed sample. It is
evident that only one mirror plane can be identified, as indicated
by the vertical line marked with $m$, implying a $2mm$ point-group
symmetry of the orthorhombic system.

In a comparative study, we have performed the same CBED study on
a $Pb-$free $Bi-$2212 single crystal and found that there are two
mirror planes in the [001] pattern, implying a $mmm$ symmetry group, which also
belongs to the orthorhombic structure. Therefore,
it seems that the $Pb-$doping lowers the symmetry of the 2212 crystal.
To date, reports on the precise symmetry of the $Bi(Pb)-$2212 system
have not reached a consensus.\cite{Fung,Goodman}
 Nevertheless, it is agreed that
the structure of the $Bi(Pb)-$2212 studied here exhibits orthorhombic, not
tetragonal, symmetry.

We also performed the same experiments on the helium annealed
$Bi(Pb)-$2212 samples. Again, no $Bi-$type modulation was observed. The
$Pb-$type modulation remained unchanged as comparing to the modulation
observed in the oxygen annealed samples. This suggests that the
change of the oxygen content does not affect the $Pb-$type modulation,
which is consistent with the results reported by Chen
et al.\cite{XHChen1,XHChen2} in their
$Pb-$doped bulk 2212 samples. No noticeable changes
were found in crystal symmetry as comparing to the
oxygen annealed samples. Only one mirror plane was observed at the [001]
CBED pattern, similar to the results for oxygen annealed sample.

\subsection{In-plane and \lowercase{c}-axis resistivity}
For lead-free $Bi-$2212, the in-plane resistivity is metallic and the
out-of-plane resistivity is highly semiconducting.\cite{Batlogg} The
resistivity anisotropy ratio of $\rho_c/\rho_{ab}$ near $T_c$ varies from
$\sim 2 \times 10^5$ (lead-free, vacuum annealed) to $\sim 800$
(lead-doped, oxygen annealed).\cite{Batlogg,Cai}.
The transport measurements on the in$-$plane and $c-$axis resistivity
were first performed on $Bi(Pb)-$2212 single crystal samples by
R\'{e}gi and co-workers.\cite{Regi} They demonstrated that
lead substitution reduced the the resistivity anisotropy
ratio $\rho_c/\rho_{ab}$ by
two orders of magnitude near the transition temperature $T_c$, compared
to the lead-free $Bi-$2212. Furthermore, they showed that the $c-$axis
resistivity
behaviour becomes metallic for lead-doped 2212.
Our resistivity measurements on $O_2$ annealed $Bi(Pb)-$2212 have fully
confirmed
their results, as shown in Fig.\ \ref{rho-c-ab}(a). A linear behavior of
$\rho_c$ occur in the temperature range of 125K$-$300K. The upward curvature
near $T_c$ suggests $c-$direction localization of carrier.

However, as shown in Fig.\ \ref{rho-c-ab}(b),
for He annealed $Bi(Pb)-$2212 we see the similar semiconducting behaviour
in the $c-$axis transport as that for lead-free 2212. X-ray diffraction,
TEM and LEED measurements exhibit no significant structural difference between
He-annealed and $O_2$-annealed lead-doped samples. These data are thus a strong
indication
that it is the amount of oxygen incorporated into a $Bi(Pb)-$2212
sample that controls the $c-$axis transport property. The presence of
lead has made it easier to get oxygen into or out of $Bi(Pb)-$2212 samples.
 As we will show
in the next section, $O_2$ doping changes both
the carrier concentration in
$Bi(Pb)-$2212 (in turn changing the chemical potential), and also the
electronic structure.

\section{Fermi Surface Topology}
The topology of the Fermi surface is an important
measure of the electronic structure. The
Fermi surface has been mapped out on lead-free
$Bi_2Sr_2CaCu_2O_{8+x}$ ($Bi-$2212) in
different oxygen-doped regimes using ARUPS by different
groups\cite{Olson1,Ma1,Aebi,Dessau1}. Unlike what has been done
for $YBa_2Cu_3O_{7-x}$ (YBCO-123)\cite{Tobin,Liu1,Liu2}, there has not yet
been a {\it systematic} study of the Fermi surface versus oxygen stoichiometry
for the $Bi-$2212 systems. We believe that the difficulty in obtaining
totally consistent results from different research groups stems from the
lack of control of oxygen stoichiometry compared to that available for the
YBCO-123 system.

Due to the comparative ease with which the oxygen stoichiometry can be
varied, the $Bi_{1.6}Pb_{0.4}Sr_2CaCu_2O_{8+x}$ ($Bi(Pb)-$2212) system
appears to be an ideal system to study the Fermi surface
topology.
Figure\ \ref{XZY} presents the complete set of
angle-resolved photoemission spectra
taken on the same oxygen overdoped $Bi_{1.6}Pb_{0.4}Sr_2CaCu_2O_{8+x}$ single
crystal sample. A photon energy of $h\nu$ = 21 eV was used. The sample was
oriented so that the $\Gamma-X$ is in the horizontal plane which is also
the photon polarization plane. The
effect of polarization on the different orientations of the sample was very
useful in determining the symmetries of the normal state bands and
will be discussed in the next section.

In Fig.\ \ref{XZY}, the spectra in each vertical column were obtained
by changing the angle
$\theta$ with fixed angle $\phi$. The spectra have been aligned horizontally
so that the spectra taken at the same $\theta$ angle, but at different $\phi$
angles are on the same horizontal level. By keeping $\theta$ fixed, one
can go along a direction parallel to the $\Gamma-Y$ symmetry direction.
Similarly, one can go along a direction parallel to the $\Gamma-X$ direction
by scanning $\theta$ with $\phi$ fixed.
We shall discuss some of the spectra in detail, below.

Using the data of Fig.\ \ref{XZY}, a purely empirical construction of the Fermi
surface of $O_2-$annealed $Bi_{1.6}Pb_{0.4}Sr_2CaCu_2O_{8+x}$ is illustrated
in Fig.\ \ref{FS}(a).
We emphasize that no {\it a priori} symmetry assumptions were made in
constructing
Fig.\ \ref{FS}(a). The Fermi surface is orthorhombic, and
$\Gamma-X$ and $\Gamma-Y$ are
inequivalent.\cite{Ma1,Osterwalder,Kelley1,Aebi}
Specifically, the shape of the Fermi surface around the $X-$point is very
similar to that reported by P. Aebi et al.\cite{Aebi} We do not have conclusive
data to comment on the presence or absence of the ``shadow bands'' reported
by P. Aebi et al.\cite{Aebi} However, again similar to
Ref.\ \onlinecite{Aebi}, the Fermi surface
nesting is not complete. Instead, the portion of the Fermi surface parallel
to $\Gamma-X$ ($k_x = k_y$) is about 60\% of the extent expected for
perfect nesting, and the nesting wavevector is approximately
$0.72(\pi,\pi)$. The portion of the
Fermi surface parallel to $\Gamma-Y$ ($k_x = - k_y$) is about 30\% of the
extent expected for perfect nesting, and the nesting wavevector is
approximately $0.80(\pi, \pi)$.

The above aspects are illustrated more explicitly in Fig.\ \ref{FS}(b).
We have assumed
only $C_{2v}$ symmetry in constructing Fig.\ \ref{FS}(b). Our nesting
wavevectors can be
directly compared to the predictions of Ruvalds et al.\cite{Ruvalds}
Ref.\ \onlinecite{Ruvalds} predicts
a marked decrease in the superconducting transition temperature ($T_c$) when
the nesting wavevector decreases from $(\pi,\pi)$. Our results agree
qualitatively with Ref.\ \onlinecite{Ruvalds}: the $T_c = 75K$ of
$O_2-$annealed $Bi(Pb)-$2212 is lower than the
$T_c = 90K$ of optimally doped $Bi-$2212. Ruvalds et al.\cite{Ruvalds}
do not obtain
quantitative agreement with our data; their predicted $T_c$ for our samples is
below 5K, compared to a measured value of 75K.

In addition, our data indicate a pocket around the $\bar{M}$ point, similar
to that
reported on $O_2-$annealed $Bi-$2212 samples by C. Olson and
colleagues.\cite{Olson1}
The portion parallel to $\Gamma-X$ (labeled ``b'' in
Fig.\ \ref{FS}(a)) merges with this
pocket, while the portion around the $X-$point (labeled ``a''
in Fig.\ \ref{FS}(a)) remains
distinct from the pocket. These results further indicate that $\Gamma-X$ and
$\Gamma-Y$ are inequivalent.

To validate the Fermi surface presented in Fig.\ \ref{FS}, we present detailed
spectra along particularly important directions.
Fig.\ \ref{GammaXYM}(a) illustrates the spectra taken along
$\Gamma-X$ direction. A band disperses towards the Fermi
energy ($E_f$)
from more than 200 meV below $E_f$ at $\theta$ of $10^\circ$ and
crosses the Fermi level at $\theta = 14^\circ$ (note abrupt
reduction of photoemission intensity). Along the $\Gamma-Y$ direction, as
illustrated in
Fig.\ \ref{GammaXYM}(b), we observe an almost equally strong band dispersing
towards $E_f$. The band
crossed the Fermi energy at $\phi=12^\circ$. The absolute positions of
crossing are slightly different with respect to the $\Gamma$ point between
$\Gamma-X$ and $\Gamma-Y$ directions, indicating that the Fermi surface
around the $X$ point might be further away from $\Gamma$ than the
Fermi surface around the $Y$ point. The difference of $2^\circ$ is right at our
combined experimental angular uncertainty.

The spectra along the $\Gamma-\bar{M}$ direction are illustrated in
Fig.\ \ref{GammaXYM}(c). A band with small dispersion is observed
 near $E_f$ and crosses the
Fermi energy at $\theta/\phi=16^\circ/16^\circ$ (note reduction in
 photoemission intensity).  We did not observe such a Fermi surface
crossing for the lead-free samples.\cite{Ma1}

The spectra indicating the turning points where the flat sections of the
Fermi surface (FS) around $X$ and $Y$ start to curve are presented
in Fig. \ \ref{P06-T12}.
In Fig.\ \ref{P06-T12}, the FS crossing happened at
$\theta=16^\circ$ for $\phi=6^\circ$, whereas the FS crossings were at
$\theta=14^\circ$ for $\phi<6^\circ$. Along the direction parallel to
$\Gamma-Y$ at $\theta=12^\circ$, the FS crossing was at $\phi=14^\circ$,
and the flat sections of FS were at $\phi=12^\circ$ with $\theta<14^\circ$.

The evidence of the existence of the pocket-like Fermi surface is shown
in Fig.\ \ref{pocket}. For a direction
parallel to $\Gamma-X$, at $\phi= 20^\circ$, a band emerges from
{\it above} the Fermi energy at about
$\theta=18^\circ$ (Fig.\ \ref{pocket}(a)), with increasing
photoemission intensity as the state disperses {\it below}
$E_f$. At $\phi= 18^\circ$, a similar Fermi surface crossing was
observed at $\theta=16^\circ$ in Fig.\ \ref{pocket}. We
already know (Fig.\ \ref{GammaXYM}(c)) that there is a FS crossing at
$\phi/\theta=16^\circ/16^\circ$. Fig.\ \ref{pocket}(c) shows two FS crossings
as one goes along a direction parallel to $\Gamma-Y$ at
$\theta=22^\circ$. The first crossing at $\phi=12^\circ$ belongs
to the FS around $X$ point. There is also a second crossing,
 at $\phi=18^\circ$, that
is part of the pocket and not part of the FS around the $Y$ point. The double
FS crossings are also present in Fig.\ \ref{pocket}(d). The first FS
crossing at $\phi=10^\circ$ corresponds to the FS around $X$ point, and
the second FS crossing at $\phi=16^\circ$ is near $\bar{M}$ but away
from $Y$ point. Finally, in Fig.\ \ref{pocket}(e), the first FS
crossing at $\phi=8^\circ$ belongs to the FS around $X$ point and
the second crossing at $\phi=16^\circ$ completes the pocket-like
Fermi surface as indicated in Fig.\ \ref{FS}. It was unfortunate that
we were not be able to take data beyond $\phi=20^\circ$ due to
technical limitations of our sample holder arrangement, but the
pocket-like FS, although incomplete, is clearly present in
the spectra in Fig.\ \ref{pocket}.

In summary, the Fermi surface map (Fig.\ \ref{FS}) and supporting data
(Figs.\ \ref{XZY},\ \ref{GammaXYM}$-$\ref{pocket})
establish several important points:
\begin{itemize}
\item the Fermi surface exhibits orthorhombic symmetry;
\item the Fermi surface nesting is quite incomplete for these HTSC's
($T_c = 75K$);
\item there is a pocket around the $\bar{M}$ point. Because we observe no
Fermi surface crossing along $\Gamma-\bar{M}-Z$ for He annealed samples,
the data
indicate that there is no pocket around the $\bar{M}$ point for He-annealed
samples. Thus, insofar as a pocket is concerned, the $Bi(Pb)-$2212
behaves in the same qualitative fashion as reported by C. Olson and
colleagues\cite{Olson1,Wells} for $Bi-$2212.
\end{itemize}

There are at least two models that view the ``flat band'' around the
$\bar{M}$ point
as important: Abrikosov's argument\cite{Abrikosov} about the effects of an
extended van
Hove singularity, and Dagotto et al.\cite{Dagotto} electron correlation
model. The
model of Dagotto et al.\cite{Dagotto} achieves quantitative agreement with
our data
on $Bi-$2212 samples.\cite{Ma1} To date, we are unaware of published
calculations
that predict the effects of a pocket on the superconducting properties. It is
clear, however, that such a fundamental change in the Fermi surface topology
can serve as a test for the above models when calculations are available. Let
us now turn to the second part of this report, the symmetry of the normal state
wavefunctions.

\section{Symmetry of the Normal States: Bloch's theorem applies}
In this section, we present data demonstrating that the symmetry of the normal
state wavefunctions changes with oxygen stoichiometry for
$Bi(Pb)-$2212 samples. Because Bloch's theorem applies, and X-ray
diffraction, TEM, and LEED measurements indicate no structural change in the
$CuO_2$ planes, there are very few plausible explanations for the data;
these are discussed after the data are presented.

We employ a standard photoemission method,\cite{Plummer} using
linearly polarized light to probe the symmetry of the normal state.
A photoemission process can be described as a transition of the
electron from the initial state $|\Psi_i>$ to a free electron final
state $|\Psi_f>$ in vacuum. The differential photoionization
cross section, to which the photoemission signal is proportional, can
be derived from the Fermi-golden-sum rule as\cite{Plummer}:
\begin{eqnarray}
\frac{d\sigma}{d\Omega}\ \propto
|<\Psi_f|\text{{\bf P}} \cdot \text{{\bf A}}_\circ|\Psi_i>|^2\
\delta(E_f-E_i-h\nu),
\end{eqnarray}
where {\bf P} is the momentum operator, {\bf A}$_\circ$
the vector potential of
photon electric field of the energy $h\nu$. We define a plane of
reflection symmetry by the $z-$axis (the surface normal) and the
photoelectron emission direction. In order to detect a
signal, the final state of the photoelectron must be even under
reflection about this symmetry plane. Otherwise, there
would be a node in the final state wave function in the
symmetry plane and no signal would be observed under any condition.
In our experimental setup, if
the {\bf A} is in the $x-$ direction, or {\bf A} $= A_x\hat{x}$, the
dipole operator {\bf P}$\cdot${\bf A} = $P_xA_x$  will
be even under the reflection about $xz-$plane ($y\rightarrow -y$), and
odd under the reflection through the $yz-$plane ($x\rightarrow -x$).
By analysing the parities of the initial state of various possible
spherical harmonics, $s$, $p$, and $d$, under the same reflection operation,
and comparing the
$<\Psi_f|\text{{\bf P}}\cdot \text{{\bf A}}|\Psi_i>$ with the signals
observed in the
experiments, we shall be able to exclude the initial states of the spherical
harmonics
that are forbidden by the dipole selection rule.\cite{Hermanson}

Figure\ \ref{O2-symmetry} illustrates normal state angle-resolved
photoemission spectra for an
$O_2-$annealed $Bi(Pb)-$2212 sample. The data include different geometries
of photon electric field vector and sample orientation, including: (a)
$\Gamma-\bar{M}-Z$ oriented in the horizontal or vertical directions;
(b) $\Gamma-X$ oriented in the horizontal or vertical directions;
(c) $\Gamma-Y$ oriented in the
horizontal or vertical directions. The insets illustrate the locations
in the Brillouin zone where the spectra were obtained. The corresponding data
for a $He-$annealed $Bi(Pb)-$2212 sample are illustrated in
Figure\ \ref{He-symmetry}. For the orientation that $\Gamma-\bar{M}-Z$ is
horizontal or vertical, the spectra are similar to
Fig.\ \ref{O2-symmetry}(a), not shown.
Consequently, we can directly compare Figs.\ \ref{O2-symmetry}(b) to
\ref{He-symmetry}(a), and
\ref{O2-symmetry}(c) to \ref{He-symmetry}(b).

Several important points emerge from the data of Figs.\ \ref{O2-symmetry}
and \ref{He-symmetry}. Both types of
samples exhibit the same symmetry (even symmetry) for the state along the
$\Gamma-\bar{M}-Z$ direction. However, there is a change of symmetry for
the state
along the $\Gamma-X$ direction. $O_2-$annealed samples exhibit a state of mixed
(even and odd) symmetry along $\Gamma-X$. By contrast, $He-$annealed
samples exhibit a state of odd symmetry along $\Gamma-X$. In addition, there is
a change of symmetry for the state along the $\Gamma-Y$ direction. For
$O_2-$annealed samples, the state along $\Gamma-Y$ exhibits odd symmetry (note
the strong dispersing peak in the vertical orientation,
Fig.\ \ref{O2-symmetry}(c)). This
strong, dispersing, peak along $\Gamma-Y$ is not observed in either orientation
for $He-$annealed samples (Fig.\ \ref{He-symmetry}(c)).

We have compared the data in Figs.\ \ref{O2-symmetry} and \ref{He-symmetry}
to the symmetry of various spherical harmonics.\cite{Plummer} A similar
analysis has been performed earlier by Kelley et al.\cite{Kelley2} for the
superconducting state of $Bi-$2212 and by Ratner et al.\cite{Ratner}
for $Pr-$doped $Bi_2Sr_2CuO_{6+y}$. We have assumed that the coupling between
the photon electric field and the near $E_f$ states along $c-$axis is so
weak that the term $P_zA_z$ in the dipole interaction can be
ignored.\cite{Plummer,Ratner} Tables\ \ref{GammaMX} and \ \ref{GammaMY}
present the comparison between experimental data of the states
along the $\Gamma-X$ , $\Gamma-Y$ and $\Gamma-\bar{M}-Z$ directions to
various spherical harmonics. For $O_2$ annealed $Bi(Pb)-$2212 samples
the symmetry of the states along $\Gamma-X$ is consistent only with either
$p_x$ or $d_{xz}$ spherical harmonic, while the states along $\Gamma-Y$ are
consistent with only a $d_{x^2-y^2}$ spherical harmonic. For $He$
annealed $Bi(Pb)-$2212 samples, the states along $\Gamma-X$ are consistent
with a $d_{x^2-y^2}$ spherical harmonic, while the states along the
$\Gamma-Y$ direction are not consistent with any single spherical harmonic.

The data of Figs.\ \ref{O2-symmetry} and \ref{He-symmetry} establish that
varying the oxygen stoichiometry
changes the symmetry of the normal state dispersing wavefunction along both
the $\Gamma-X$ and $\Gamma-Y$ directions. The data were taken in the normal
state, so within a one electron picture Bloch's theorem applies. Consequently,
a change in symmetry is due to a change in the spatial orientation of the
wavefunction. We have looked for, and found no indication of, a change in
the structure of the $CuO_2$ planes. We have, however, observed a change in
the interlayer coupling (Fig.\ \ref{rho-c-ab}). In the absence of a
structural change in the $CuO_2$ planes, the data compel us to conclude that
the change in wavefunction symmetry is caused by the change in interlayer
coupling.

The above conclusion is within the context of a one-electron picture, for
which Bloch's theorem holds. However, as we mention earlier,
Dagotto et al.\cite{Dagotto} recently obtained
quantitative agreement with our data\cite{Ma1} for the $Bi-$2212 system.
Ref.\ \onlinecite{Dagotto} argues that the dispersion relations arise
from many-body effects, within the context of a $t-J$ model. Calculations
are in progress to extend these results by including next-nearest-neighbor
interactions.\cite{Dagotto2} At
present, we are unaware of any available many-body calculation to which we
can compare our data.

\section{B\lowercase{i}(5\lowercase{d}) and
P\lowercase{b}(5\lowercase{d}) Core Levels}
As noted earlier, one motivation for our study was the earlier results of
C. Olson and colleagues,\cite{Olson1} which did not appear consistent
with a rigid
band picture. Other investigators\cite{Pham} have also argued that a rigid band
picture is inappropriate for the cuprate superconductors. The change in
wavefunction symmetry along $\Gamma-X$ and $\Gamma-Y$, as concluded in
the previous section, are also
inconsistent with a two-dimensional, rigid-band, picture. We used another
method to determine whether a rigid band model is appropriate. In the
rigid band model, changing the oxygen stoichiometry (the carrier
concentration) will change the chemical potential. Such a change of chemical
potential would produce a rigid shift of all electronic states, including
the valence band and core levels.

We thus measured the valence band and $Bi(5d)$ core levels for three types of
samples: as-grown $Bi-$2212 ($T_c = 90K$), $O_2-$annealed
$Bi(Pb)-$2212 ($T_c = 75K$), and $He-$annealed $Bi(Pb)-$2212
($T_c = 85K$). Fig.\ \ref{Bi-Pb-5d} illustrates the results.
The data in Fig.\ \ref{Bi-Pb-5d} are inconsistent
with a rigid band model. We first used the as-grown $Bi-$2212 samples
as a reference. The conduction band of the $O_2-$annealed $Bi(Pb)-$2212
samples were shifted by 310 meV to lower binding energy. However, the $Bi(5d)$
core levels of the same samples were shifted by only 60 meV to lower binding
energy. For the $He-$annealed $Bi(Pb)-$2212 samples, the conduction
band was shifted by 160 meV to lower binding energy, while the $Bi(5d)$ core
levels were shifted by 250 meV to higher binding energy. Thus, the rigid
band model, which would result in the same shift of all states-
is inconsistent with the data.

In addition, the rigid band model is also inconsistent with the data taken
from only lead-doped samples. As Fig.\ \ref{Bi-Pb-5d} illustrates, we also
measured the
$Pb(5d)$ core level for such samples. Using $O_2-$annealed
$Bi_{1.6}Pb_{0.4}Sr_2CaCu_2O_{8+x}$
samples as a reference, the shifts of the He-annealed
$Bi_{1.6}Pb_{0.4}Sr_2CaCu_2O_{8+x}$
samples include the valence band (150 meV higher binding energy), the $Bi(5d)$
core level (310 meV higher binding energy), and the $Pb(5d)$ core level
(260 meV higher binding energy).

In summary, we do observe a shift of the chemical potential to lower absolute
energy for $O_2-$annealed samples, as expected. However, the rigid band model
is not consistent with our data.

\section{Conclusions}

{}From our TEM data,
we have demonstrated that $Bi_{1.6}Pb_{0.4}Sr_2CaCu_2O_{8+x}$ is
structurally orthorhombic and its crystal structure is not sensitive to
oxygen stoichiometry we applied. However,
we have observed a large reduction in $c-$axis resistivity for $O_2-$annealed
$Bi_{1.6}Pb_{0.4}Sr_2CaCu_2O_{8+x}$ samples, consistent with earlier
reports. The Fermi
surface of such samples exhibits orthorhombic symmetry, incomplete Fermi
surface nesting, and a pocket around the $\bar{M}$ point (which seems absent
for He-annealed samples).

In addition, there is a change in the symmetry of the normal state dispersing
band along both the $\Gamma-X$ and $\Gamma-Y$ directions with oxygen
stoichiometry.
Within the one-electron band structure picture, the data indicate that the
change in symmetry is caused by the change in interlayer coupling as
indicated by the change of $c-$axis resistivity.

By measuring the shift in both the chemical potential (valence band) and
core levels, we have established that the rigid band picture does not apply
to the $Bi_{1.6}Pb_{0.4}Sr_2CaCu_2O_{8+x}$ system, consistent with earlier
reports on other cuprate systems.

\vspace{.4in}
\noindent ACKNOWLEDGEMENTS

We benefitted from conversations with E. Dagotto and results prior to
publication. C. Quitmann kindly provided technical assistance.
We acknowledge the financial support provided by the U.S.
NSF, both directly (DMR-9214707)  and through support of the SRC, by Ecole
Polytechnique F\'{e}d\'{e}rale Lausanne and the Fonds National Suisse de la
Recherche Scientifique.
\eject

\begin{figure}
\caption{The [001](a) and [100](b) electron diffraction patterns (EDP) of
the oxygen annealed $Bi(Pb)-$2212 single crystal samples. Diffraction
spots are indexed based on the
fundamental structure. Satellite spots, as indicated by arrows, are due
to the modulated structure. Similar EDP were observed on the helium annealed
$Bi(Pb)-$2212 single crystal samples.}
\label{EDP}
\end{figure}

\begin{figure}
\caption{[001]Convergent beam electron diffraction (CBED) pattern of the oxygen
annealed $Bi(Pb)-$2212 single crystal samples. The pattern has only one
mirror plane as indicated by the vertical line marked $m$, implying
a 2mm point-group symmetry of orthorhombic structure. A similar
CBED pattern was observed on the helium annealed $Bi(Pb)-$2212
single crystal samples.}
\label{CBED}
\end{figure}

\begin{figure}
\caption{Temperature dependence of the $c-$axis resistivity and in-plane
resistivity for $Bi_{1.6}Pb_{0.4}Sr_2CaCu_2O_{8+x}$ single crystals (a)
annealed
in 1 atmosphere $O_2$ at 600 $^\circ$C for 1 hour; (b) annealed in 1
atmosphere $helium$ at 600 $^\circ$C for 1 hour.}
\label{rho-c-ab}
\end{figure}

\begin{figure}
\caption{The normal state (T = 95K) angle-resolved photoemission
spectra (EDC) measured on a oxygen overdoped
$Bi_{1.6}Pb_{0.4}Sr_2CaCu_2O_{8+x}$ single crystal sample
of $T_c$ = 75K using
photon energy $h\nu$ = 21 eV. The photoelectron emission angle relative
to the surface normal ($\theta$ and $\phi$) are marked. The spectra in
each vertical column were taken by changing the angle $\theta$ with fixed
angle $\phi$. The data were taken in the
range $0 \leq \phi \leq 20^\circ$ and $0 \leq \theta \leq 40^\circ$ which
covers almost 2/3 of the first Brillouin zone.}
\label{XZY}
\end{figure}

\begin{figure}
\caption{(a). The experimentally determined Fermi surface of $Bi(Pb)-$2212
from EDC measurements using $h\nu$ = 21 eV. The data were taken on 160
points in the first Brillouin zone. No symmetry assumption is used.
All of data points are experimental ones. (b). The Fermi surface of
$Bi(Pb)-$2212
constructed out of experimental data points and assumption of $C_{2v}$
symmetry.}
\label{FS}
\end{figure}

\begin{figure}
\caption{Normal state (T = 95K) angle-resolved photoemission spectra for
an oxygen overdoped $Bi_{1.6}Pb_{0.4}Sr_2CaCu_2O_{8+x}$ single crystal
of $T_c$ = 75K along (a) the $\Gamma-X$ direction; (b) the $\Gamma-Y$
direction; (c) the $\Gamma-\bar{M}-Z$ direction. The photon energy employed
was 21 eV. The insets show the locations in the first Brillouin zone where
the data were taken.}
\label{GammaXYM}
\end{figure}

\begin{figure}
\caption{Normal state (T = 95K) angle-resolved photoemission spectra for
an oxygen overdoped $Bi_{1.6}Pb_{0.4}Sr_2CaCu_2O_{8+x}$ single crystal
of $T_c$ = 75K along (a) a direction parallel to $\Gamma-X$ at
$\phi=6^\circ$, where the Fermi surface crossing happens at
$\theta=16^\circ$ instead of $\theta=14^\circ$ for $\phi < 6^\circ$;
(b) a direction parallel to $\Gamma-Y$ at $\theta=12^\circ$, where the
Fermi Surface crossing is at $\phi=14^\circ$ instead of $\phi=12^\circ$
for $\theta < 12^\circ$.}
\label{P06-T12}
\end{figure}

\begin{figure}
\caption{Normal state (T = 95K) angle-resolved photoemission spectra for
an oxygen overdoped $Bi_{1.6}Pb_{0.4}Sr_2CaCu_2O_{8+x}$ single crystal of
$T_c$ = 75K along the directions parallel to $\Gamma-X$ (a) at
$\phi=20^\circ$; (b) at $\phi= 18^\circ$, and the directions parallel to
$\Gamma-Y$ (c) at $\theta=22^\circ$; (d) at $\theta=20^\circ$; and (e) at
$\theta=18^\circ$. The pocket-like Fermi surface around $\bar{M}$ is
derived from those cuts.}
\label{pocket}
\end{figure}

\begin{figure}
\caption{Normal state (T= 95K) angle-resolved photoemission spectra
for an $O_2$ annealed $Bi_{1.6}Pb_{0.4}Sr_2CaCu_2O_{8+x}$ single
crystal of $T_c$ = 75K for different geometries of photon
polarization vector and crystalline orientation.
(a) $\Gamma-\bar{M}-Z$ is horizontal or vertical; (b) $\Gamma-X$ is horizontal
or vertical; (c) $\Gamma-Y$ is horizontal or vertical. The photon
polarization vector is always in the horizontal plane. The insets
show the locations in the Brillouin zone where the spectra were obtained.}
\label{O2-symmetry}
\end{figure}

\begin{figure}
\caption{Normal state (T = 95K) angle-resolved photoemission spectra for
a {\it helium annealed} $Bi_{1.6}Pb_{0.4}Sr_2CaCu_2O_{8-x}$
single crystal of $T_c$ = 85K for different geometries of photon polarization
vector and crystalline orientation. For the orientation that
$\Gamma-\bar{M}-Z$ is horizontal or vertical, the spectra are similar
to Fig. 9(a). However, the signal observed are quite different for the
orientations
(a) $\Gamma-X$ is horizontal or
vertical; (b) $\Gamma-Y$ is horizontal or vertical. The insets show
the locations in the BZ where the spectra were taken.}
\label{He-symmetry}
\end{figure}

\begin{figure}
\caption{(a).The $Bi(5d)$ core levels for
as-grown $Bi-$2212 ($T_c$ = 90K), The $Bi(5d)$ and
$Pb(5d)$ core levels for $O_2$ annealed $Bi(Pb)-$2212 ($T_c$ = 75K), and
He annealed
$Bi(Pb)-$2212 ($T_c$ = 85K). The binding energies are labeled nearby the core
level peaks. The
photon energy $h\nu$ is 50 eV. (b). The valence spectra for undoped $Bi-$2212
($T_c$ = 90K) and $O_2$ annealed $Bi(Pb)-$2212 ($T_c$ = 75K) and He annealed
$Bi(Pb)-$2212 ($T_c$ = 85K).
Notice the shifts of main feature which consists of mostly Cu(3d) valence.
The photon energy $h\nu$ = 21 eV is used. We emphasize that the core levels
and valence spectra were taken on the same sample of different materials.
Only the photon energy of the monochromator are different.}
\label{Bi-Pb-5d}
\end{figure}

\begin{table}
\caption{A list of possible $s$, $p$, or $d$ symmetries of the normal state
with orientations for which a normal state band along $\Gamma-\bar{M}-Z$
and $\Gamma-X$ could be (Yes) or could not (No), by symmetry , be observed.}
\begin{tabular}{ccccc}
 & \multicolumn{2}{c}{$\Gamma-\bar{M}-Z$}
& \multicolumn{2}{c}{$\Gamma-X$}\\
Orientation   & horizontal & vertical & horizontal & vertical\\
Operation     & $y\rightarrow -\ y$           & $y\rightarrow -\ y$
                 & $x\leftrightarrow -\ y$          & $x\leftrightarrow -\ y$\\
Symmetry        & \                             & \
                 & \                             & \ \\
\tableline
$s$      & $<\! e|e|e\! >$\ Yes          & $<\! e|o|e\! >$\ No\
         & $<\! e|e|e\! >$\ Yes         & $<\! e|o|e\! >$\ No\ \\
$p_x$    & $<\! e|e|e\! >$\ Yes          & $<\! e|o|e\! >$\ No\
         & $<\! e|e|e\! +\! o\! >$\ Yes  & $<\! e|o|e\! +\! o\! >$\ Yes\\
$p_y$    & $<\! e|e|o\! >$\ No           & $<\! e|o|o\! >$\ Yes
         & $<\! e|e|e\! +\! o\! >$\ Yes  & $<\! e|o|e\! +\! o\! >$\ Yes\\
$p_z$          & $<\! e|e|e\! >$\ Yes         & $<\! e|o|e\! >$\ No\
               & $<\! e|e|e\! >$\ Yes         & $<\! e|o|e\! >$\ No\ \\
$d_{xy}$       & $<\! e|e|o\! >$\ No\         & $<\! e|o|o\! >$\ Yes
               & $<\! e|e|e\! >$\ Yes         & $<\! e|o|e\! >$\ No\ \\
$d_{x^2-y^2}$  & $<\! e|e|e\! >$\ Yes         & $<\! e|o|e\! >$\ No\
               & $<\! e|e|o\! >$\ No\         & $<\! e|o|o\! >$\ Yes\\
$d_{3z^2-r^2}$ & $<\! e|e|e\! >$\ Yes         & $<\! e|o|e\! >$\ No\
               & $<\! e|e|e\! >$\ Yes    & $<\! e|o|e\! >$\ No\ \\
$d_{yz}$ & $<\! e|e|o\! >$\ No\          & $<\! e|o|o\! >$\ Yes
         & $<\! e|e|e\! +\! o\! >$\ Yes  & $<\! e|o|e\! +\! o\! >$\ Yes\\
$d_{xz}$ & $<\! e|e|e\! >$\ Yes          & $<\! e|o|e\! >$\ No\
         & $<\! e|e|e\! +\! o\! >$\ Yes  & $<\! e|o|e\! +\! o\! >$\ Yes\\
\tableline
Experiment & \ & \ & \  & \ \\
$O_2$ $Bi(Pb)-$2212 & Yes & No\  & Yes & Yes\\
$He$ $Bi(Pb)-$2212 & Yes & No\  & No\ & Yes\\
\end{tabular}
\label{GammaMX}
\end{table}

\begin{table}
\caption{A list of possible $s$, $p$, or $d$ symmetries of the normal state
with orientations for which a normal state band along $\Gamma-\bar{M}-Z$
and $\Gamma-Y$ could be (Yes) or could not (No), by symmetry , be observed.}
\begin{tabular}{ccccc}
 & \multicolumn{2}{c}{$\Gamma-\bar{M}-Z$}
 & \multicolumn{2}{c}{$\Gamma-Y$}\\
Orientation & Horizontal &Vertical & Horizontal & Vertical\\
Operation  & $y\rightarrow -\ y$           & $y\rightarrow -\ y$
                 & $x\leftrightarrow \ y$          & $x\leftrightarrow \ y$\\
Symmetry        & \                             & \
                 & \                             & \ \\
\tableline
$s$             & $<\! e|e|e\! >$\ Yes           & $<\! e|o|e\! >$\ No\
                 & $<\! e|e|e\! >$\ Yes           & $<\! e|o|e\! >$\ No\ \\
$p_x$        & $<\! e|e|e\! >$\ Yes           & $<\! e|o|e\! >$\ No\
          & $<\! e|e|e\! +\! o\! >$\ Yes & $<\! e|o|e\! +\! o\! >$\ Yes\\
$p_y$  & $<\! e|e|o\! >$\ No         & $<\! e|o|o\! >$\ Yes
       & $<\! e|e|e\! +\! o\! >$\ Yes    & $<\! e|o|e\! +\! o\! >$\ Yes\\
$p_z$           & $<\! e|e|e\! >$\ Yes      & $<\! e|o|e\! >$\ No\
                 & $<\! e|e|e\! >$\ Yes     & $<\! e|o|e\! >$\ No\ \\
$d_{xy}$        & $<\! e|e|o\! >$\ No\      & $<\! e|o|o\! >$\ Yes
                 & $<\! e|e|e\! >$\ Yes    & $<\! e|o|e\! >$\ No\ \\
$d_{x^2-y^2}$   & $<\! e|e|e\! >$\ Yes      & $<\! e|o|e\! >$\ No\
                 & $<\! e|e|o\! >$\ No\     & $<\! e|o|o\! >$\ Yes\\
$d_{3z^2-r^2}$  & $<\! e|e|e\! >$\ Yes      & $<\! e|o|e\! >$\ No\
                 & $<\! e|e|e\! >$\ Yes     & $<\! e|o|e\! >$\ No\ \\
$d_{yz}$        & $<\! e|e|o\! >$\ No\      & $<\! e|o|o\! >$\ Yes
          & $<\! e|e|e\! +\! o\! >$\ Yes    & $<\! e|o|e\! +\! o\! >$\ Yes\\
$d_{xz}$        & $<\! e|e|e\! >$\ Yes          & $<\! e|o|e\! >$\ No\
                 & $<\! e|e|e\! +\! o\! >$\ Yes & $<e|o|e\! +\! o\!>$\ Yes\\
\tableline
Experiment & \  & \  & \  & \ \\
$O_2$ $Bi(Pb)-$2212 & Yes & No\  & Yes & No\ \\
$He$ $Bi(Pb)-$2212  & Yes & No\  & No\  & No\ \\
\end{tabular}
\label{GammaMY}
\end{table}

\end{document}